\documentclass[12pt,preprint]{aastex}
\usepackage{emulateapj5}
\begin{document}
\newcommand{\myemail}{ian.george@gsfc.nasa.gov}
\shorttitle{X-rays from NGC~3226}
\shortauthors{George et al.}
\journalinfo{Accepted for publication in ApJ}
\title{The X-ray Emission from the Nucleus of the 
Dwarf Elliptical Galaxy NGC~3226}

\author {I.M. George\altaffilmark{1,2},
R.F. Mushotzky\altaffilmark{1},
T. Yaqoob\altaffilmark{1,3},
T.J. Turner\altaffilmark{1,2},
S. Kraemer\altaffilmark{4,5},
A.F. Ptak\altaffilmark{6},
K. Nandra\altaffilmark{1,7},
D.M. Crenshaw\altaffilmark{4,5},
H. Netzer\altaffilmark{8}
}

\altaffiltext{1}{Laboratory for High Energy Astrophysics, Code 662,
        NASA/Goddard Space Flight Center,
        Greenbelt, MD 20771}
\altaffiltext{2}{Joint Center for Astrophysics, 
        Department of Physics,
	University of Maryland, Baltimore County, 
	1000 Hilltop Circle, Baltimore, MD 21250}
\altaffiltext{3}{Department of Physics \& Astronomy, 
	Johns Hopkins University, 3400 North Charles Street, 
	Baltimore, MD 21218}
\altaffiltext{4}{Laboratory for Astronomy and Solar Physics, Code 681,
         NASA/Goddard Space Flight Center,
        Greenbelt, MD 20771}
\altaffiltext{5}{Institute for Astrophysics and Computational Sciences, 
        The Catholic University of America, Washington D.C. 20064}
\altaffiltext{6}{Department of Physics, Carnegie Mellon University, 
Pittsburgh, PA 15213}
\altaffiltext{7}{Universities Space Research Association}
\altaffiltext{8}{School of Physics and Astronomy and the Wise Observatory,
        The Beverly and Raymond Sackler Faculty of Exact Sciences,
        Tel Aviv University, Tel Aviv 69978, Israel.}

\slugcomment{Accepted for publication in {\em The Astrophysical Journal}}

\begin{abstract}
We present the first high resolution X-ray image of the dwarf elliptical 
galaxy NGC~3226. The data were obtained during an observation of the 
nearby Seyfert Galaxy NGC~3227 using the {\it Chandra X-ray Observatory}.
We detect a point X-ray source spatially consistent with the optical 
nucleus of NGC~3226 and a recently-detected, compact, flat-spectrum, 
radio source. The X-ray spectrum can be measured up to $\sim$10~keV
and is consistent with a power law with a photon index
$1.7 \lesssim \Gamma \lesssim 2.2$, or thermal bremmstrahlung emission 
with $4 \lesssim kT \lesssim 10$~keV. In both cases the luminosity in the 
2--10~keV band $\simeq 10^{40} h_{75}^{-1}\ {\rm erg\ s^{-1}}$.
We find marginal evidence that the nucleus varies within the observation. 
These characteristics support evidence from other wavebands that 
NGC~3226 harbors a low-luminosity, active nucleus. 
We also comment on two previously-unknown, fainter X-ray sources 
$\lesssim 15$~arcsec from the nucleus of NGC~3226. Their proximity to the 
nucleus (with projected distances $\lesssim 1.3 h_{75}^{-1}$~kpc) 
suggests both are within NGC~3226, and thus have luminosities 
($\sim$few$\times10^{38}$--few$\times10^{39}\ {\rm erg\ s^{-1}}$)
consistent with black-hole binary systems.
\end{abstract}


\section{INTRODUCTION}
\label{sec:intro}

\noindent
NGC~3226 is a dwarf elliptical galaxy
($z=0.00441\pm0.00006$; e.g. de Vaucouleurs et al. 1991)
interacting with the nearby SAB(s)a galaxy NGC~3227\footnote{For 
simplicity, here we use the heliocentric velocity  
($1322\pm19\ {\rm km\ s^{-1}}$) measured using optical emission lines
to derive the distance to NGC~3226. 
Given its interaction with NGC~3227 ($1157\pm3\ {\rm km\ s^{-1}}$), 
the distance may be slightly over-estimated (by $\lesssim14$\%), 
but this does not affect any of the conclusions presented here.}.
The center of NGC~3226 hosts a low-ionization nuclear
emission-line region (LINER) of optical spectroscopic type 1.9 
(Ho, Fillipenko \& Sargent, 1997a), along with large quantities of dust 
in filamentary structures (e.g. Rest et al. 2001).
A compact, flat-spectrum radio source has recently been detected 
(Nagar et al. 2000; Falcke et al. 2000) at a location consistent 
with the optical nucleus. This, along with the claim of a broad component 
to the H$\alpha$ emission line (FWHM $\sim 2\times10^{3}\ {\rm km\ s^{-1}}$; 
Ho et al. 1997b), suggests the galaxy hosts an active galactic 
nucleus (AGN) rather than a region of aging starburst activity.
The luminosity of the broad H$\alpha$ line in NGC~3226
($L$(H$\alpha$) $\simeq10^{39}\ {\rm erg\ s^{-1}}$) is 
an order of magnitude below the dividing line between so-called 
low-luminosity AGN (LLAGN, or ``dwarf'' Seyfert galaxies) and the 
more powerful (``normal'') Seyfert 1 and giant elliptical galaxies.

X-ray emission from the nuclear regions of NGC~3226 has been detected 
using both the {\it ROSAT}\ PSPC (Komossa \& Fink 1997; 
Sansom, Hibbard \& Schweizer 2000) and HRI (Roberts \& Warwick 2000)
with an implied luminosity $L$(0.1--2~keV) $\sim 10^{40}\ {\rm erg\ s^{-1}}$.
However these data sets had relatively poor spatial resolution, 
and were unable to distinguish either spatially or spectroscopically
between an (AGN-like) power law X-ray spectrum and a
collisionally ionized plasma due to hot gas 
(e.g. from a starburst region).

In this {\it Letter} we present results for NGC~3226 obtained using 
the {\it Chandra X-ray Observatory} 
({\it CXO}: e.g. Weisskopf, O'Dell, van Speybroeck 1996).
Following a brief description of the 
observation (\S\ref{Sec:Obs}), in \S\ref{Sec:spatial} we report the 
detection of three X-ray sources within the inner $\sim 2$~kpc of the 
nucleus, with brightest source identified as the nucleus of 
NGC~3226. In \S\ref{Sec:3226} we report the temporal and 
spectral characteristics of the nucleus, and discuss the 
evidence that NGC~3226 harbors a LLAGN. The limited constraints that can be 
obtained from the two off-nuclear sources are discussed in 
\S\ref{Sec:spec-others}. We present our conclusions in \S\ref{Sec:conclusions}.

\begin{figure*}[t]
\includegraphics[scale=0.5,angle=0]{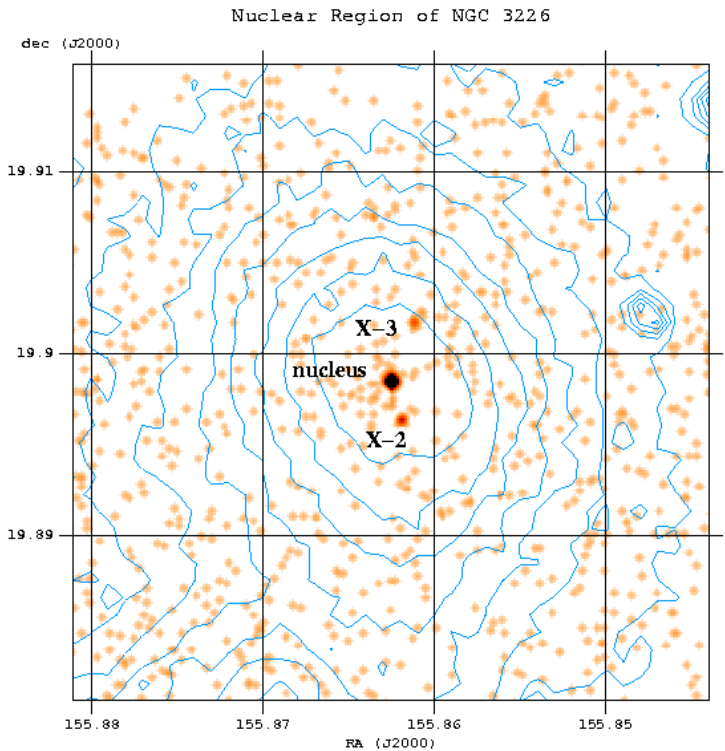}
\caption{The $0^{\rm th}$-order HETGS image of the nuclear region of 
NGC~3226 in the 0.8-6~keV band. 
For clarity the image has been smoothed by a Gaussian 
with $\sigma = 0.492$~arcsec. 
The DSS image is overlaid 
as the blue, linearly-spaced contours.
The separation between the grid lines ($10^{-2}$~degrees) 
$\simeq 3.1 h_{75}^{-1}$~kpc.
The offsets between the position of the brightest X-ray source  
and the optical nucleus of NGC~3226 as tabulated by 
Cotton, Condon \& Arbizzani (1999) are
($\Delta$RA, $\Delta$dec) $=$ ($-1.2$~arcsec, $+0.2$~arcsec).
Currently there are no optical identifications of the other 
two X-ray sources (marked X-2 \& X-3).
\label{fig:spatial}}
\end{figure*}

\section{THE OBSERVATION}
\label{Sec:Obs}

\noindent
The data from NGC~3226 reported here were obtained during an observation
using the {\it CXO} in 1999 Dec.
The High Energy Transmission Grating Spectrometer 
(HETGS: e.g. Markert et al. 1994) was employed with the 
Advanced CCD Imaging Spectrometer (ACIS: e.g. Nousek et al 1998)
in the focal plane (with a temperature of $-110^{\circ}$~C).
The HETGS consists of two sets of gratings (hereafter the 
medium- and high-energy grating \{MEG \& HEG\} ``arms''), 
each with a different period and intercepting X-rays from 
(two of the four) different shells of the X-ray telescope.
The rulings of the MEG and HEG differ by $\sim$10~degrees
preventing any spatial overlap in the dispersed spectra of 
an on-axis, point source.
The ACIS-S array was placed at the optical axis, consisting 
of 6 CCD chips\footnote{The ACIS-S0 chip was not in operation during the
observation due to instrumental concerns.}
orientated to record the undispersed ($0^{\rm th}$-order) 
photons from the target and nearby sources (on the ACIS-S3 CCD),
along with both the positive and negative orders of the spectrum 
dispersed by both grating arms of the HETGS.

The observation was continuous, with a total duration of 
49.9~ks. Following screening and removing the ``streaks'' from 
the s4 chip using {\tt destreak} (v~1.3; Houck 2000), 
the resultant exposure time was 46.3~ks.
All the data analysis presented here was performed using
the {\it Chandra} {\tt CALDB} (v2.0), and the 
{\tt CIAO} (v2.0.2) and {\tt HEAsoft} (v5.0.2) software packages.
Further details on the 
instrumental parameters and the data reduction
can be found in George et al. (2001), 
along with the scientific results for NGC~3227.

\section{SPATIAL ANALYSIS}
\label{Sec:spatial}

\noindent
In Fig.~\ref{fig:spatial} we show an image of NGC~3226 in the 0.6-8~keV band.
For simplicity, here we ignore any peculiar motion associated with the 
NGC~3226/3227 system or other members of its local group, 
and simply use the heliocentric velocity of  NGC~3226 
to derive a  distance to $17.6 h_{75}^{-1}$~Mpc 
(where $h_{75} = H_0/75\ {\rm km\ s^{-1}\ Mpc^{-1}}$).  
Thus 1~arcsec $\equiv 85 h_{75}^{-1}$~pc. 
Three X-ray sources are evident. 
We find the centroid of the brightest source to be at 
RA=10h~23m~27.00s, dec=+19d~53m~54.7s (J2000).
This is consistent the position of the optical nucleus of NGC~3226 
(RA=10h~23m~27.08s, dec=+19d~53m~54.5s; Cotton, Condon \& Arbizzani 1999)
within the current uncertainties in the aspect reconstruction
of {\it CXO} data ($\lesssim$1~arcsec; Aldcroft, p.comm),
and in excellent agreement with the compact, flat-spectrum radio source 
(RA=10h~23m~27.01s, dec=+19d~53m~54.5s; Falcke et al. 2000).
Thus we identify this X-ray emission as the HETGS $0^{\rm th}$-order
image of the inner $\sim150 h_{75}^{-1}$~pc of NGC~3226.
From an analysis using a Marr wavelet function provided by the {\tt wavdetect} 
task in {\tt CIAO} we estimate a net source count rate of 
$491\pm22\ {\rm count\ s^{-1}}$.

To the best of our knowledge, nothing has been reported in the literature 
concerning the other two sources, CXO~J102334.1+195347 and 
CXO~J102326.7+195407 (labelled X-2 \& X-3 respectively in 
Fig.~\ref{fig:spatial}). Both are much weaker than the nucleus 
(with estimated source count rates of $20\pm5\ {\rm count\ s^{-1}}$
and $18\pm4\ {\rm count\ s^{-1}}$ respectively).

\section{THE NUCLEUS OF NGC~3226}
\label{Sec:3226}

\begin{figure*}
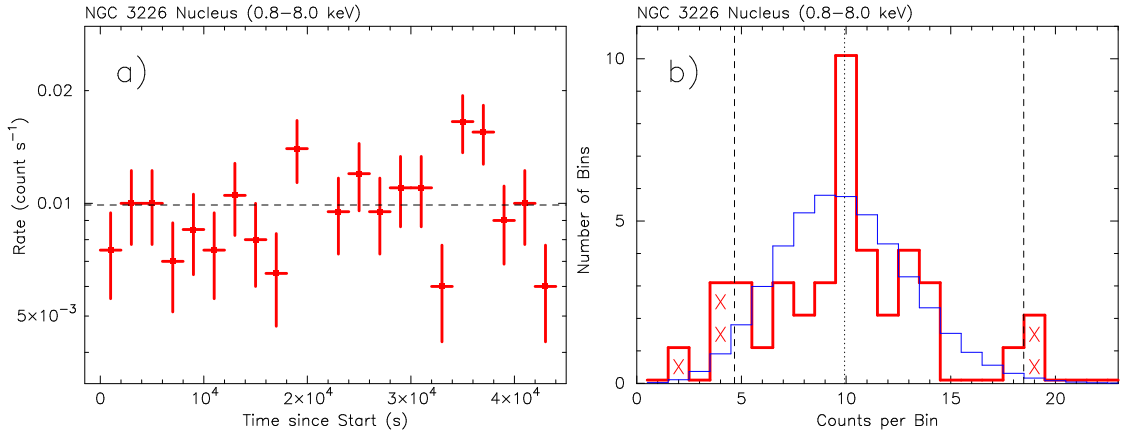

\includegraphics[scale=0.3,angle=270]{imgeorge_fig2a.ps}
\includegraphics[scale=0.3,angle=270]{imgeorge_fig2b.ps}
\caption{a) A light curve of the nucleus of NGC~3226
in the 0.8--6~keV band using $2\times10^3$~s bins.
Note that the y-axis is logarithmic, with a dynamic range 
0.33-3.0 times the mean count rate ($0.0099\ {\rm ct\ s^{-1}}$;
dashed line).
The predicted background rate is 
$\simeq 3\times10^{-5}\ {\rm count\ s^{-1}}$
and constant. 
b) Observed distribution (red) of number of bins 
(each of $10^3$~s) containing 
$n$ counts, and a Poisson distribution (blue)
derived from the mean count rate. 
The dashed lines show the 95\% confidence limits for the 
Poisson distribution. The bins marked with an ``X'' highlight 
more bins are observed outside these limits than predicted.
\label{fig:lc1000}}
\end{figure*}

\noindent
For the (temporal and spectral analysis) presented below we 
employed a circular extraction cell of radius $\simeq2.5$~arcsec
centered on the nucleus. The background was obtained from a
circular annulus with an outer radius $\simeq25$~arcsec centered 
on the nucleus, but also excluding circular regions of radius 
$\simeq2.5$~arcsec around X-2 and X-3. The background was then 
rescaled to be appropriate for the source extraction cell.

\subsection{Temporal Variability}
\label{Sec:3226-temp}

\noindent
We find marginal evidence for variability from the nucleus of NGC~3226.
In Fig.~\ref{fig:lc1000}a we show the observed count rate 
in 0.8--6~keV band as a function of time using $2\times10^3$~s bins.
There are 21 such bins, each containing $\sim$20 counts on average, 
thus $\chi^2$-statistics can be used (but with some caution).
The hypothesis of a constant count rate is rejected at $\sim94$\% 
confidence ($\chi^2_{\nu}= 1.53$ for 20 degrees of freedom, 
hereafter d.o.f.).
Using $4\times10^3$~s bins, a constant count rate is rejected at 
$\sim97$\% confidence ($\chi^2_{\nu}= 1.96$ for 10 d.o.f.).

Given the data from NGC~3226 are in the ``grey-area'' where 
the appropriateness of using $\chi^2$-statistics is debatable
(i.e. $\lesssim 20$ bins, each with a relatively small number of 
counts) we have also performed a temporal analysis assuming 
Poissonian statistics. This analysis also indicates evidence 
for variability. For example,  
in Fig.~\ref{fig:lc1000}b we show the observed 
(red) number of bins (each of $10^3$~s), 
containing 0, 1, 2, .... counts compared to the predicted 
Poisson distribution (blue) derived from the mean count rate of 
$\simeq 0.01\ {\rm ct\ s^{-1}}$.  
For the example shown, we find 6 of the 42 bins outside the 
the 95\% confidence limits (dashed lines in Fig.~\ref{fig:lc1000}b)
compared to the 2.1 predicted from a Poisson distribution.
We have repeated this analysis 20 times, on each occasion 
offsetting the start time of each $10^3$~s bin by 1/20 of a bin 
width. We find the observed number of bins exceeds the predicted 
number on 13 of the 20 occasions. 
Similarly we have repeated the analysis using 
bins of different temporal sizes, and find cases where number of bins observed 
outside the 95\% confidence limits exceeds that expected 
for bins in the $\sim$few$\times10^{2}$--few$\times10^{3}$~s.
Nevertheless, given the current data are in the Poissonian regime, 
again we consider the evidence only marginal.

\subsection{Spectral Analysis}
\label{Sec:3226-spec}

\begin{figure*}
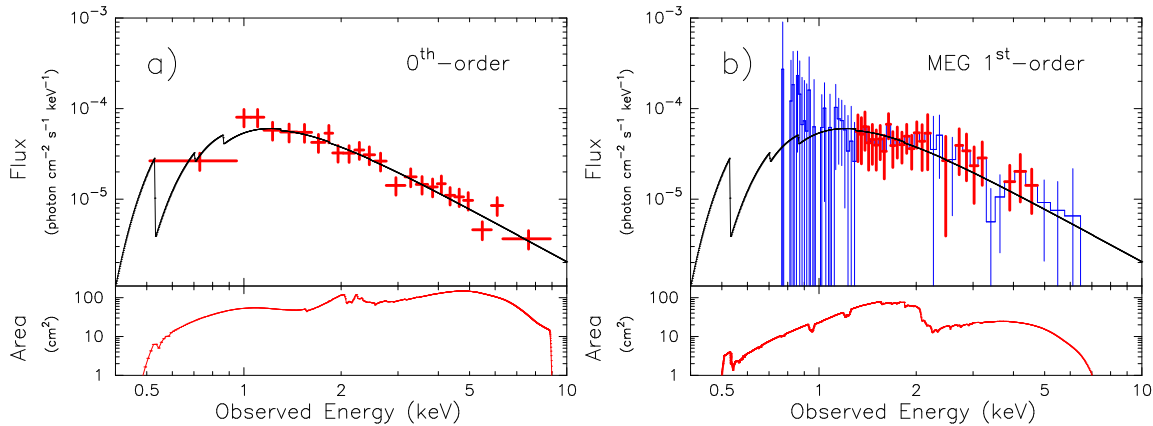

\includegraphics[scale=0.3,angle=270]{imgeorge_fig3a.ps}
\includegraphics[scale=0.3,angle=270]{imgeorge_fig3b.ps}
\caption{The upper panels show the spectra from the nucleus of NGC~3226,
along with the absorbed, power-law model discussed in 
the text. Bins which contain more than 1 photon at
$> 97.5$\% confidence are shown in red.
The lower panels we show the effective area of the 
instruments.
a) The $0^{\rm th}$-order spectrum 
grouped such that each bin contains $\geq 20$ counts.
b) The MEG $1^{\rm st}$-order spectrum derived using 
only data $<1.6$~nm ($-1^{\rm st}$-order)
and $<1.8$~nm ($+1^{\rm st}$-order)
to maximize the signal-to-noise ratio.
For clarity the data are shown using 
23~pm wide bins (ie. $10\times$ the FWHM of the MEG spectral resolution).
No statistically significant features are present in the spectrum 
made at the full MEG resolution.
\label{fig:nuc_smth_spec}}
\end{figure*}

\noindent
Spectra were extracted from the ($0^{\rm th}$-order) cell
described above using the {\tt psextract} 
script in {\tt CIAO}. The resultant Pulse-Invariant (PI) 
channels at energies $>0.5$~keV were then grouped so as to contain a 
minimum of 20 counts per bin, and hence allowing 
a $\chi^{2}$ minimization technique to be employed.
This resulted in 24 spectral bins which were analyzed 
within the {\tt XSPEC} (v 11.0.1) package.
For the Galactic absorption along the line-of-sight
we assume an effective hydrogen column density of
$N_{H\,I}^{gal} = 2.08\times10^{20}\ {\rm cm^{-2}}$ 
(as measured towards NGC~3227; Murphy et al 1996).

We find an absorbed power law provides an adequate fit to the data
($\chi^{2} = 20.1$ for 21 degrees of freedom).
The best-fitting parameters and 68\% confidence ranges are 
a photon index $\Gamma = 1.94^{+0.26}_{-0.25}$, and 
intrinsic absorption (in addition to $N_{H\,I}^{gal}$) of 
$N_{H} = (4.8^{+1.7}_{-1.5})\times10^{21}\ {\rm cm^{-2}}$.
The implied luminosity (corrected for absorption) in the 
0.4--10~keV band is $L$(0.4--10~keV) 
  $\simeq 3.5^{+1.4}_{-1.0}\times10^{40} h_{75}^{-1}\ {\rm erg\ s^{-1}}$,
and that in the 2--10~keV band $L$(2--10~keV) 
  $\simeq 1.8^{+0.8}_{-0.3}\times10^{40} h_{75}^{-1}\ {\rm erg\ s^{-1}}$.
The data and this model are shown in Fig.\ref{fig:nuc_smth_spec}a.
Unfortunately any Fe K$\alpha$ emission cannot be 
well constrained by the data
(equivalent widths $EW \lesssim 0.9$~keV and 
$\lesssim 1.5$~keV for narrow and broad lines at 6.4~keV
respectively).
An absorbed thermal bremsstrahlung also provides an adequate 
fit to the data 
($\chi^{2} = 22.2$ for 21 degrees of freedom), 
with $kT = 6.5^{+4.4}_{-2.1}$~keV, 
and with a luminosity and $N_{H}$ 
(in addition to $N_{H\,I}^{gal}$) consistent with those found above.
We shall discuss these two models further in \S\ref{Sec:discuss}.

The nucleus of NGC~3226 is sufficiently bright to also be 
detected in the dispersed spectrum. Source and background 
spectra were extracted using the procedure described in George et al. (2001). 
The source exceeds the mean background level in the 
$\pm1^{\rm st}$-orders in the ranges $0.18$--$1.7$~nm (MEG)
and $0.15$--$1.0$~nm (HEG). 
The spectral resolution of the MEG and HEG can be well 
approximated by Gaussians with (``$1 \sigma$'') widths 
$\simeq 1$~pm and $\simeq 0.5$~pm respectively
($1$~pm $\equiv 10^{-12}$~m; and also corresponds to 
$\sim$1 ACIS pixel\footnote{Due to the orientation of the grating arms 
    and  
    the dithering of the spacecraft during the observation
    the relationship between ACIS pixel and wavelength is 
    complex and time-variable.}).
Both the MEG and HEG spectra  have a very low signal-to-noise ratio 
($\lesssim 0.5\ {\rm ct\ pm^{-1}}$ and $\lesssim 0.3\ {\rm ct\ pm^{-1}}$ 
respectively). We have constructed spectra using bins between 1 and 4 times 
the resolution of each grating arm.
 However we find no cases where the number of counts 
per bin exceeds 7 ($\simeq 99$\% confidence limit for 1 count per bin),
and hence no evidence that the spectrum contains any intense emission 
lines. Using larger bin sizes, we find both the MEG and HEG spectra are 
in good agreement with that derived from the $0^{\rm th}$-order
(e.g. Fig.\ref{fig:nuc_smth_spec}b).
Thus we find no requirement for additional spectral complexity
from either the $0^{\rm th}$-order or dispersed spectra.
 
Our spectral models predict {\it ROSAT} HRI count rates 
($\sim4\times10^{-3}\ {\rm ct\ s^{-1}}$)
consistent with those seen during an observation in 
1995 Aug (Roberts \& Warwick 2000).
However the predicted {\it ROSAT} PSPC count rates 
($\sim10^{-2}\ {\rm ct\ s^{-1}}$)
are slightly lower than that reported during 
an observation in 1993 May 
($\sim 4\times10^{-2}\ {\rm ct\ s^{-1}}$
from within a circular region of radius 
$\sim15 h_{75}^{-1}$~kpc is reported by Sansom et al. 2000).
This may be the result of variability on long timescales, and/or
an additional soft X-ray component from the 
AGN. However at least part of the higher count rate seen by the 
PSPC may be due to other sources of (perhaps time-variable) 
X-ray emission associated with the galaxy.

\subsection{The Case for a LLAGN at the Center of NGC~3226}
\label{Sec:discuss}

\noindent
The improved spatial resolution of {\it CXO} has enabled us to 
show that the X-ray emission from NGC~3226 is dominated by a point-like 
(tentatively variable) source within $\sim0.2 h_{75}^{-1}$~kpc of the nucleus.
The X-ray spectrum is measured up to $\sim$10~keV
and is consistent with a power law with a photon index
$1.7 \lesssim \Gamma \lesssim 2.2$, or thermal bremsstrahlung emission 
with $4 \lesssim kT \lesssim 10$~keV.
The ratio of observed flux at 2~keV ($\simeq 4\times10^{-2}\ \mu$Jy)
to that at 5~GHz from the flat-spectrum, compact radio source
(Falcke et al. 2000) 
gives a radio-to-X-ray spectral index 
$\alpha_{rx} = - \log (f({\rm 2~keV})/f({\rm 5~GHz}))/7.986 = 0.62$.
Such a value is consistent with that derived from several similar 
objects, including M~81 (NGC~3031) and NGC~4579 
(both with $\alpha_{rx} = \simeq 0.6$; e.g. Ho 1999, and references therein).
M~81 and NGC~4579 are of note since both  
are commonly believed to be accretion-powered objects. 
Indeed Fe K$\alpha$ emission has been detected from both 
objects, which is broad in the case of M~81 (Ishisaki et al. 1996)
and variable in the case of NGC~4579 (Terashima et al. 2000), 
highly suggestive of an accretion disk.

In common with many other LLAGN, optical images reveal 
large quantities of dust in filamentary structures 
within the inner regions of NGC~3226 (e.g. Rest et al. 2001).
The X-ray presented here also indicate intrinsic absorption. 
The equivalent hydrogen column density reported in 
\S\ref{Sec:3226-spec} (3--$7\times10^{21}\ {\rm cm^{-2}}$)
corresponds to $2 \lesssim A_{V} \lesssim 4$
assuming a Galactic dust-to-gas ratio.
Using the spectral energy distributions of 
M~81 and NGC~4579 as templates, both of which have 
an optical-to-X-ray spectral index
$\alpha_{ox} = -  \log (f({\rm 2~keV})/f(250~{\rm nm}))/2.605 \simeq 1.0$
(after correcting for the better-known reddening in these galaxies),
we predict the nucleus has $m_V \simeq m_R \gtrsim 21$. 
The nucleus is therefore predicted to be swamped by the stellar emission
(Rest et al. find $\mu_R \sim 14\ {\rm arcsec^{-2}}$) in the central 
regions, consistent with observations.

These characteristics strongly suggest NGC~3226 hosts a 
central AGN.
Such an hypothesis is also consistent with the broad component to 
the H$\alpha$ emission line suggested by Ho et al. (1997b)
based on modeling of a fairly noisy spectrum of the 
H$\alpha$ region (where H$\alpha$ is blended with 
\verb+[+N {\sc ii}\verb+]+ and \verb+[+S {\sc ii}\verb+]+;
see their Fig. 10a).
Terashima, Ho \& Ptak (2000) have reported a 
strong correlation between $L$(2--10~keV) and the
luminosity of H$\alpha$ for a number of galaxies containing 
type 1 LINERs. This correlation has recently been confirmed and extended
by Ho et al. (2001). 
This correlation supports the hypothesis 
that the dominant ionization source of
the type 1 LINERs is the photon field of a central LLAGN.
NGC~3226 is consistent with this correlation, adding 
further support to this hypothesis.
The fact that the observed broad component of H$\alpha$ 
constitutes a lower faction ($\sim60$\%) of the 
total (broad plus narrow) line emission compared to 
a more ``normal'' Seyfert 1 galaxy (eg. $\sim97$\% in 
NGC~3516) suggests differential (probably patchy) absorption 
towards the narrow- and broad-line regions. 

\subsubsection{Constraints on the Accretion Process}

\noindent
The optical luminosity of the entire galaxy NGC~3226
($M_{\rm B_T} = -19.4$; $L_{\rm B}/L_{\odot} \simeq 8\times10^{9}$)
and a central, line-of-sight velocity dispersion 
($\simeq 180\ {\rm km\ s^{-1}}$; Simien \& Prugniel 1998)
are consistent with the relation found for a 
variety of galaxies harboring black holes
(eg Gebhardt et al. 2000a; and references therein).
The implied central 
black hole mass is $\simeq 10^{8}\ M_{\odot}$, and hence 
the implied accretion rate is 
{\it \.{m}} $\sim 4\times10^{-6}  h_{75}^{-1} f_{\rm bolX}$ 
{\it \.{M}}$_{\rm Edd}$, 
where $f_{\rm bolX}$ is the ratio of bolometric to 0.4--10~keV 
luminosity (which is unlikely to much exceed 10), 
and 
{\it \.{M}}$_{\rm Edd}$ is the Eddington accretion rate.
Some words of caution are necessary. 
First, we note that the line-of-sight velocity dispersion quoted 
above is derived from ground-based observations -- as yet there 
have been no {\it HST}\ observations capable of resolving the 
central kinematics of NGC~3226.
Second, there have recently been a number of questions 
raised regarding the exact relation between the luminosity 
of the bulge and the mass of the black hole estimated 
by various means in different samples 
(eg. see Gebhardt et al. 2000b; Merritt \& Ferrarese 2001, 
and references therein).
Nevertheless, if a black hole indeed exists at the center of 
NGC~3226, the accretion rate is surely 
{\it \.{m}}$ \lesssim 10^{-2}${\it \.{M}}$_{\rm Edd}$.

In systems with such low rates, the 
mass infall is generally thought to occur via
a ``standard'' (geometrically-thin, viscosity-dominated) 
accretion disk at large radii, but switch to a radiatively 
less-efficient advection-dominated accretion flow 
(ADAF) in the inner regions 
(e.g. Narayan, Mahadevan \& Quataert 1998, and references therein).
The transition radius at which the flow becomes 
advection-dominated depends on a number of model parameters.
Estimates range from $\sim10$--$10^2$ Schwarzschild radii 
for several LLAGN (including M~81 and NGC~4579; 
e.g. see Quataert et al. 1999) 
to $\sim10^3$ Schwarzschild radii for giant elliptical galaxies 
within clusters (Di Matteo et al. 2000).
The X-ray emission from ADAFs is believed to be 
dominated by thermal bremsstrahlung (with $kT \sim$few keV), 
and not expected to exhibit significantly variability on timescales 
shorter than the local dynamical time 
($\sim 5\times10^{4}$~s at 10 Schwarzschild radii 
from a $10^{8}\ M_{\odot}$ black hole -- see also Ptak et al. 1998).
In \S\ref{Sec:3226-spec} we found a thermal bremsstrahlung
model to be consistent with the observed spectrum of the nucleus.
However, if the variability on 
timescales $\lesssim$few$\times10^{3}$~s suggested in \S\ref{Sec:3226-temp} 
is confirmed, time-dependent models will be required.

\section{THE OFF-NUCLEAR SOURCES CXO~J102334.1+195347 AND CXO~J102326.7+195407}
\label{Sec:spec-others}

\noindent
There are too few counts from either 
CXO~J102334.1+195347 AND CXO~J102326.7+195407
to enable detailed analysis. We find 
no evidence for statistically significant variability
on any timescale.
Nevertheless some crude insight can be gained from consideration 
of counts detected in different energy bands.
For CXO~J102334.1+195347, the ratio of source counts 
(0.3-2~keV)/(2--10~keV) $=0.84\pm0.40$. 
For a power law spectrum, such a ratio can be obtained with 
$0.9 \lesssim \Gamma \lesssim 1.8$ if absorption by only 
$N_{H\,I}^{gal}$ is assumed, 
or with $\Gamma = 2$ and additional absorption 
of $N_{H}\simeq 10^{21}$--$10^{22}\ {\rm cm^{-2}}$.
In the case of CXO~J102326.7+195407, we find the ratio of counts 
(0.3-2~keV)/(2--10~keV) $\lesssim 0.2$, indicative of an even harder 
spectrum.
Assuming absorption by only $N_{H\,I}^{gal}$, a power law with 
$\Gamma \lesssim 0.5$ is required.
Assuming a power law with $\Gamma = 2$, additional 
absorption with $N_{H}\gtrsim 2\times10^{22}\ {\rm cm^{-2}}$
is required.

The proximity of both sources to the nucleus 
(within a projected distance $\sim 1.3 h_{75}^{-1}$~kpc)
strongly suggests both are X-ray sources within NGC~3226.
The observed flux from each source is 
$\sim$few$\times10^{-14}\ {\rm erg\ cm^{-2}\ s^{-1}}$
in the 2--10~keV band.
The density of background sources is $\lesssim 10^{2}\ {\rm deg^{-2}}$
in this band at this flux level
(e.g. Mushotzky et al 2000; Tozzi et al. 2001; and references therein).
Thus the probability of a background source this bright 
within 15~arcsec of the nucleus of NGC~3226 is $\lesssim 10^{-4}$.

Assuming both sources are indeed within NGC~3226, the spectral forms 
described above imply absorption-corrected luminosities
in the range 
 $\sim$few$\times10^{38}$ 
  -- few$\times10^{39} h_{75}^{-1}\ {\rm erg\ s^{-1}}$.
Assuming all emission from each source is indeed due to a 
single object, either or both CXO~J102334.1+195347 
and CXO~J102326.7+195407 therefore have luminosities 
exceeding the Eddington limit for neutron stars 
($\sim 3\times10^{38}\ {\rm erg\ s^{-1}}$), and 
hence must be $\gtrsim$few$M_{\odot}$ black-holes, 
very young supernovae or microquasars.
Such superluminal sources are now being found regularly 
by {\it CXO}\ (eg. Griffiths et al. 2000; 
Fabbiano, Zezas \& Murray 2001).
Variability and detailed X-ray spectroscopic studies are the most 
likely means the various source types can be distinguished.

\section{CONCLUSIONS AND FUTURE PROSPECTS}
\label{Sec:conclusions}

\noindent
The results presented here illustrate the power of 
X-ray observations to detect and characterize the emission from 
LLAGN: a class of source which can usually only otherwise be studied
at radio wavelengths and possibly
with high signal-to-noise, high-resolution 
optical emission-line studies.
Indeed in many cases, as for NGC~3226, the nucleus is expected to be 
invisible to {\it HST} in the optical band.
We anticipate that many such objects will be detected serendipitously 
in the fields of future {\it CXO} and {\it XMM-Newton} 
observations.
These data will expand the regions of parameter space 
accessible to study, surely resulting in a better understanding of 
the AGN/black-hole phenomenon.

\acknowledgements

We thank Lorella Angelini, Mike Loewenstein 
and Raymond White~III for useful discussions.
We acknowledge the financial support of the
Joint Center for Astrophysics (IMG, TJT), 
NASA (TJT, through grant number NAG5-7385 (LTSA)),
and
the Universities Space Research Association (KN).
This research has made use of the Simbad database, operated by 
the Centre de Donn\'{e}es astronimiques de Strasbourg (CDS);
the {\tt VizieR} Service for Astronomical Catalogues,
developed by CDS and ESA/ESRIN;
the Hypercat Extragalactic database, provided by the
Centre de Recherche Astronomique de Lyon;
the NASA/IPAC
Extragalactic Database (NED), operated by the Jet Propulsion Laboratory,
California Institute of Technology, under contract with 
NASA; 
and of data obtained through the HEASARC on-line service, 
provided by NASA/GSFC.


\newpage
\typeout{REFERENCES}

\end{document}